\documentclass[a4paper]{article}

\usepackage{INTERSPEECH2020}

\usepackage{float}

\title{Competitive Wakeup Scheme for Distributed Devices}
\name{Lu Ma, Haiping Zhang, Pei Zhao, Tengrong Su}
\address{Haier Smart Home Co., Ltd.}
\email{malu@haier.com, iamroad@163.com}

\begin{document}

\maketitle
\begin{abstract}
  Wakeup is the primary function in voice interaction which is the mainstream scheme in man-machine interaction (HMI) applications for smart home. All devices will response if the same wake-up word is used for all devices. This will bring chaos and reduce user quality of experience (QoE). The only way to solve this problem is to make all the devices in the same wireless local area network (WLAN) competing to wake-up based on the same scoring rule. The one closest to the user would be selected for response. To this end, a competitive wakeup scheme is proposed in this paper with elaborately designed calibration method for receiving energy of microphones. Moreover, the user orientation is assisted to determine the optimal device. Experiments reveal the feasibility and validity of this scheme.
\end{abstract}

\section{Introduction}
Voice interaction is the mainstream scheme in man-machine interaction (HMI) scenarios, especially for smart home \cite{interaction}. Among the techniques referred to in voice interaction, wakeup is the primary function \cite{wakeup}. All devices will response if the same wake-up word is used for all devices. This will give sense of chaos for users, thus reduce user quality of experience (QoE). The only way to address this problem is to make all the devices in the same wireless local area network (WLAN) \cite{WLAN} competing to wake-up. The one closest to the user would be selected for response. To this end, a fair scoring rule is needed for judging which device is the one closest to the user.

The frequently utilized quantity to calculate distance is arriving time \cite{TOA}, that is, multiple devices can obtain the signal arrival time and send this time value to the central node through the network to make the decision where the one corresponding to the earliest arrival time is selected as the responsor. This requires that the devices could obtain accurate time information in real-time, and time synchronization could be guaranteed among these devices. However, latency and instability would be encounted by network in practice.  Moreover, sine different signal processing algorithm may be adopted for each device, an additional unknown delay would be introduced, which could not be obtained accurately through testing. These all completely deny the time-based scheme. Due to the fact that household devices would always be stationary, infrequent movement or tiny movement, it is impracticable to employ the variation of sound frequency which is wideband covering from a low frequency to a high one for judgments \cite{doppler}. So the frequency-based scheme is infeasible. Therefore, only the quantity of magnitude or energy can be used for scoring \cite{RSSI, RSSI1}.

When employing the quantity of energy for scoring, parameter calibration must be performed among these devices, especially for the microphone gain, that is, all microphones are normalized to the same standard microphone \cite{array}. This is also the current mainstream calibration scheme which is accomplished artificially in advance. However, due to the hardware differences of household devices and the poor consistency of microphones, it is impossible to tune the parameters in advance with the increasing device. This gives the demand of a user self-calibration scheme which is elaborately designed in this paper, that is, turn the microphone calibration work to the user by adding the calibration function at the APP of mobile phone. All the devices in the same WLAN can be configured through APP by the user itself. Aside from the calibration of microphone gain, the interference introduced by speakers of devices during play is also eliminated for accurate energy calculation aimed for deciding which device for response. To improve the user sense, the user orientation is considered for better determining the optimal device to response.

\section{Competitive Wakeup Scheme}
\subsection{System architecture of competitive wakeup}
\begin{figure}
\centering
\includegraphics[scale=0.4]{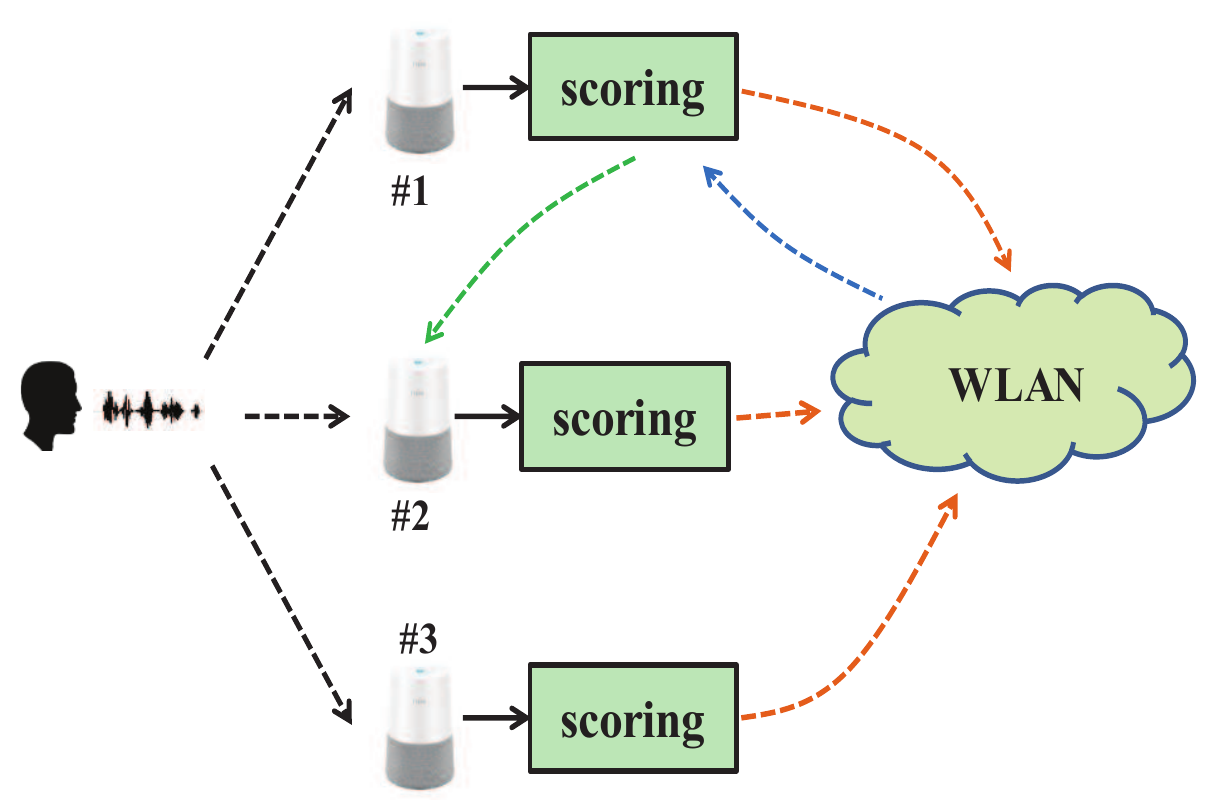}
\caption{System deployment}
\label{fig:deployment}
\end{figure}

The structure of competitive wakeup is depicted in Fig. \ref{fig:deployment} where three devices are considered as an example. When wakeup word is yelled out by user, all the devices in the same WLAN will receive the audio signal. In each device, scoring algorithm is called to calculate the decision quantity, this is, the energy of wakeup word in this scheme. Then, this energy value is sent to the master device which is selected based on the network quality of the device through WLAN. All these energy values are calibrated and then compared to find the largest one in the master device. Finally the device labeled $\#2$ considered closest to the user is selected to response while the others keep silence.

\begin{figure}
\centering
\includegraphics[scale=0.5]{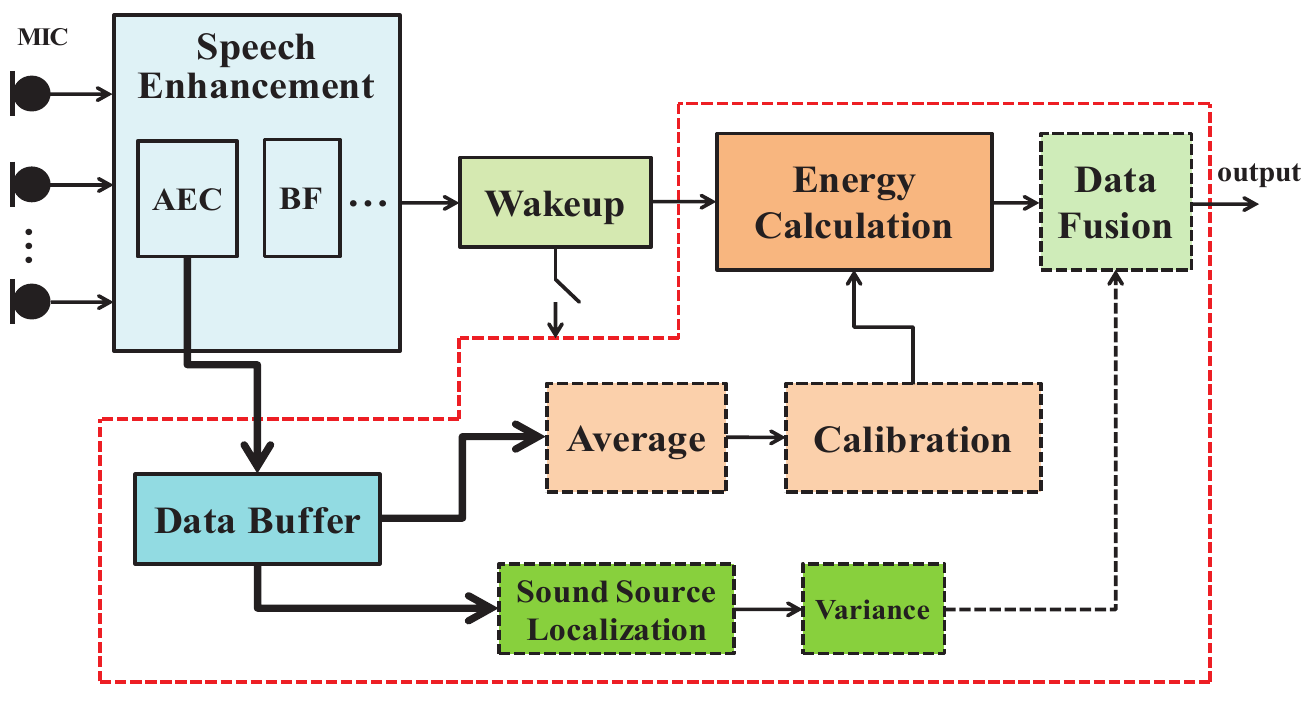}
\caption{Structure of scoring algorithm}
\label{fig:scoring}
\end{figure}

The scoring algorithm called in each device is illustrated in Fig. \ref{fig:scoring}. After receiving the wakeup word, speech enhancement such acoustic echo cancellation (AEC), array beamforming (BF), noise suppression (NS) is first implemented to obtain a superior audio signal \cite{enhance}. This signal is then calculated by wakeup algorithm. Only when wakeup is detected, a copy of multi-channel data after AEC is reserved at the data buffer. These data are then used for energy calculation and sound source localization (SSL). Before energy calculation, these multi-channel data are superposed and averaged to one channel, then calibration is done to normalize the microphone gain of the device itself and eliminate the speaker interference that may be received from other devices. Thereafter, a pure energy of wakeup word is calculated and used for determining distance with respect to the user. Meanwhile, a series of direction-of-arrival (DOA) for each frame are calculated based on the buffered data, and then the variance of these DOAs are obtained. This is used for determining the user face orientation. The energy and the orientation are used together to determine the optimal device for response.

\subsection{Calibration Method}
For obtaining a pure energy of wakeup word at the same level of hardware for all devices, microphone gain should be calibrated and normalized to the same standard high-performance microphone which is prepared in advance. In addition to this, the audio interference of speakers that may come from other devices should also be eliminated for accurate energy calculation. Here, these calibration methods will be illustrated in detail from the perspective of APP end and device end respectively.

\subsubsection{Gain calibration of microphone}
The calibration for the device microphone is conducted by normalizing the receiving energy to the standard one measured in advance by high-performance microphone. When the microphone calibration is performed by the APP, it will prompt the user to stand at the front of the device with a distance set by the APP. In this circumstance, the APP will play the audio of the wakeup word stored in the cellphone, and the device will calculate the energy by the scoring algorithm after receiving the audio signal, which is denoted by ${E_k}_{,{d_i}}$, where $d_i$ is the distance, $k$ is the device ID, the subscript $i$ corresponds to the distance index. This energy will be transmitted back to the APP where normalization to the standard energy denoted by ${E_0}_{,{d_i}}$ calculated from the standard high-performance microphone at that distance will be realized, obtaining the coefficient as
\begin{equation}
{b_k}_{,{d_i}} = \frac{{{E_{k,{d_i}}}}}{{{E_0}_{,{d_i}}}}
  \label{eq1}
\end{equation}

\begin{figure}
\centering
\includegraphics[scale=0.45]{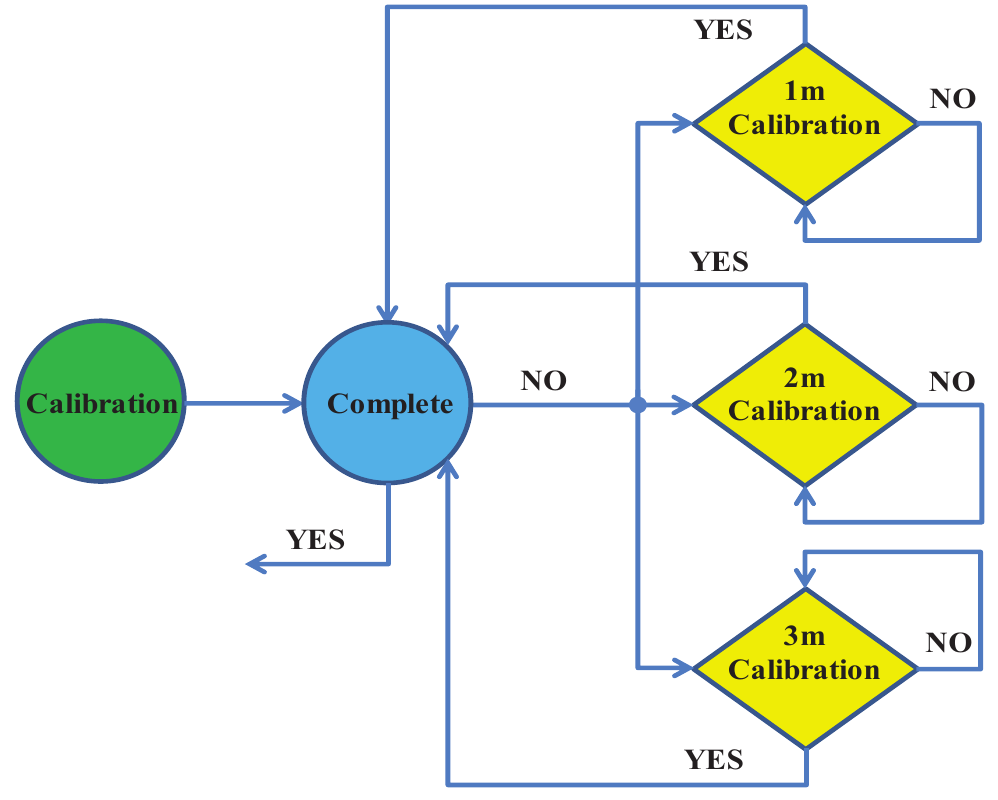}
\caption{Logical flowchart of microphone calibration}
\label{fig:gainlogic}
\end{figure}

Once the coefficient corresponding to that distance is obtained, the audio playing of the wakeup word is stopped. Calibration at different distance should be accommodated to obtain higher accuracy while considering the room layout (for example, the room space is small resulting in infeasible for 3m calibration or the device is suspended on the high resulting in infeasible for 1m calibration). After calibrations at all the distances are achieved, the calibration of microphone gain is finished, and the coefficients calculated at these distances will be weighted averaged as the final calibration coefficient. The flowchart of the logical control for calibration is shown in Fig. \ref{fig:gainlogic}.

\subsubsection{Interference elimination of speaker}
Only when the microphone gain calibration is finished, could the interference elimination of speaker be performed. When the device plays audios, there is little interference for the device itself due to the AEC algorithm based on the reference channel. However, it will bring interference to other devices since no reference signal of this interference are obtained in these devices, affecting the energy calculation for fair competition.

\begin{figure}[H]
\centering
\includegraphics[scale=0.5]{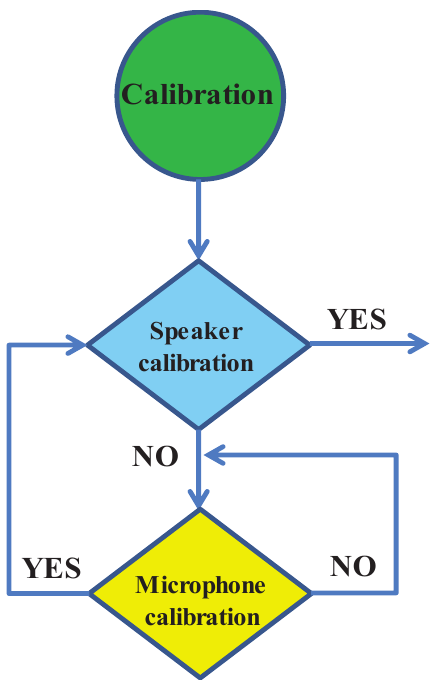}
\caption{Logical flowchart for interference elimination}
\label{fig:speakerlogic}
\end{figure}

Fortunately, this can be done by making one of the devices playing wakeup word while the others keep listening and calculating the interference energy which is obtained on listening devices. This manipulation is implemented among the devices one by one until all the interferences from one device to the others are calculated. The logical control of interference elimination of speaker is depicted in Fig. \ref{fig:speakerlogic}.

Suppose that the device labeled $i$ is playing wakeup word, the energy of the reference channel denoted by $E_i$ will be calculated with energy calculation algorithm at the device. At the same time, the device labeled $j$ have also received this audio signal, and the audio energy is calculated as the interference denoted by $E_{i,j}$. In the same way, the interference energy with respect to other devices can be obtained. When all the interference energy are calculated with respect to device $i$, the wakeup word playing for device $i$ is stopped. In the same way, the next device will be controlled playing wakeup word, and the interference energy of other devices with respect to this device is calculated and transmitted to the APP. When all the interference energy is calculated, an interference matrix would be obtained for interference elimination.

For example, when device 1 is for elimination, the APP controls the device playing wakeup word. In this time, the reference energy of device 1 is calculated and transmitted back to the APP. As long as three consecutive handshakes are accomplished, the reference energy of device 1 denoted by $E_{1,1}$ is obtained by averaging these three values. Meanwhile, the interference from device 1 to device 2, device 1 to device 3 and device 1 to device 4 can be calculated in the same way and are denoted by $E_{1,2}$, $E_{1,3}$,  $E_{1,4}$. As long as all the interference from device 1 to others and the reference energy of itself are obtained at the APP, could the coefficients be calculated by
\begin{equation}
{a_1} = \left[ {\begin{array}{*{20}{c}}
{1,}&{a{}_{1,2}}&{{a_{1,3}}}&{{a_{1,4}}}
\end{array}} \right]\begin{array}{*{20}{c}}
,
\end{array}\begin{array}{*{20}{c}}
{}
\end{array}{a_{i,j}} = \frac{{{E_{i,j}}}}{{{E_i}}}
  \label{eq2}
\end{equation}
where ${a_{1,2}} = \frac{{{E_{1,2}}}}{{{E_1}}}$, ${a_{1,3}} = \frac{{{E_{1,3}}}}{{{E_1}}}$, ${a_{1,4}} = \frac{{{E_{1,4}}}}{{{E_1}}}$, and the interference coefficients from other devices can be obtained in the same way, thus getting the following coefficient matrix as
\begin{equation}
A = \left[ {\begin{array}{*{20}{c}}
1&{{a_{1,2}}}&{{a_{1,3}}}&{{a_{1,4}}}\\
{{a_{2,1}}}&1&{{a_{2,3}}}&{{a_{2,4}}}\\
{{a_{3,1}}}&{{a_{3,2}}}&1&{{a_{3,4}}}\\
{{a_{4,1}}}&{{a_{4,2}}}&{{a_{4,3}}}&1
\end{array}} \right]
  \label{eq3}
\end{equation}

For concise expression, all the coefficients on the diagonal can be replaced by the coefficients of the calibrated microphone gain denoted by $b_i$ (where $i$ is the device ID, such as: 1, 2, 3, 4), therefore the calibration matrix can be expressed by
\begin{equation}
A = \left[ {\begin{array}{*{20}{c}}
{{b_1}}&{{a_{1,2}}}&{{a_{1,3}}}&{{a_{1,4}}}\\
{{a_{2,1}}}&{{b_2}}&{{a_{2,3}}}&{{a_{2,4}}}\\
{{a_{3,1}}}&{{a_{3,2}}}&{{b_3}}&{{a_{3,4}}}\\
{{a_{4,1}}}&{{a_{4,2}}}&{{a_{4,3}}}&{{b_4}}
\end{array}} \right]
  \label{eq4}
\end{equation}

\subsection{User orientation calculation}
The direct parameter for determining the voice orientation is the multi-path information, i.e., the energy ratio between the direct path and the reflected ones. The user is more likely to be considered as front orientation if this ratio is lager, and it is more likely to be at side orientation if the ratio is smaller. However, the energy of the direct path and the reflected ones are difficult to be obtained directly.

\begin{figure}[H]
\centering
\includegraphics[scale=0.55]{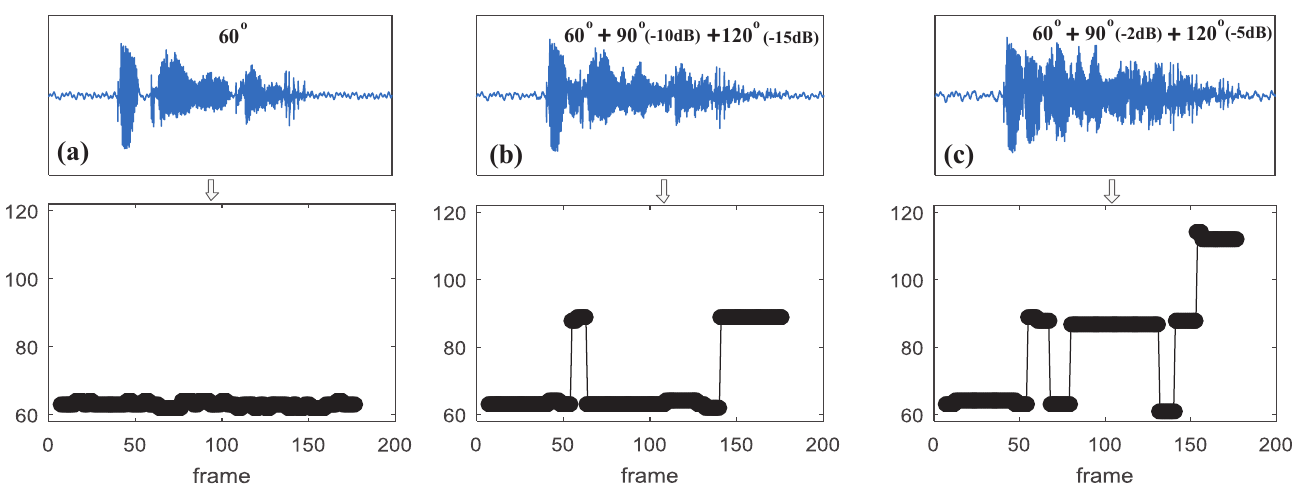}
\caption{User orientation calculation based on DOA variance}
\label{fig:doa}
\end{figure}

Fortunately, experiments found that DOA precision can be affected by multi-path. This can be depicted in Fig. \ref{fig:doa} where three scenarios are evaluated, i.e., a) only direct path at $60^\circ$ with 0dB energy, b) one $60^\circ$ direct path superposed by one $90^\circ$ reflection with -10dB energy and one $120^\circ$ reflection with -15dB energy, c) one $60^\circ$ direct path superposed by one $90^\circ$ reflection with -2dB energy and one $120^\circ$ reflection with -5dB energy. The user orientation diverges gradually from a) to c), and the variance increases from $0.4^\circ$ to $129.8^\circ$ and $279.6^\circ$. This reveals that the DOA variance for consecutive frames increases with the orientation deviation, i.e., the DOA variance formulated as follows can be used for determining user orientation.

\begin{equation}
G_i = \frac{1}{K}{\sum\limits_{k = 1}^K {\left( {{\varphi _k} - \frac{1}{K}\sum\limits_{k = 1}^K {{\varphi _k}} } \right)} ^2}
  \label{eq6}
\end{equation}
where $G_i$ is the localization variance of the $i$-th device, $\varphi _k$ is the DOA result of the $k$-th frame, and $K$ is the total number of frames for calculation.

\section{Scoring Algorithm}
Herein, energy calculation of wakeup word is elaborately designed to improve the accuracy and robust to the environment. The energy values of all the devices will be calibrated and then sorted to find the one with maximal energy as responsor to user.

\subsection{Energy calculation}
\begin{figure}[H]
\centering
\includegraphics[scale=0.45]{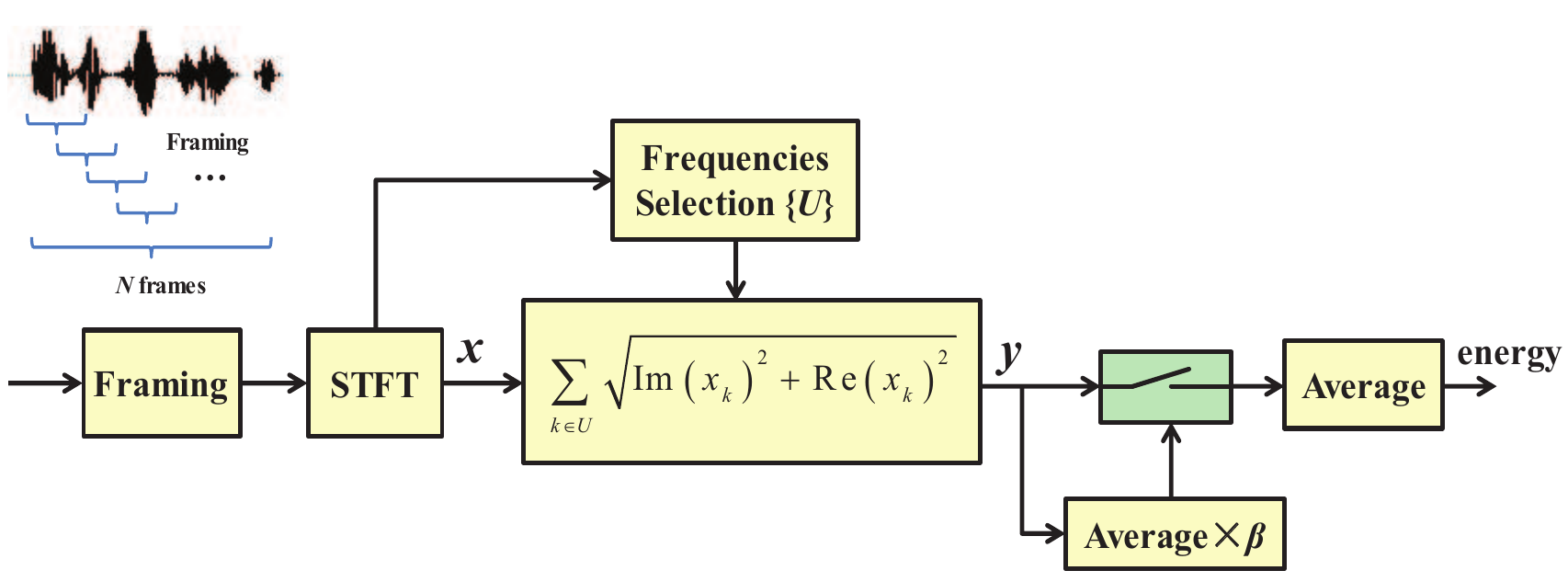}
\caption{Energy calculation at device}
\label{fig:energy}
\end{figure}

The energy calculation is illustrated in Fig. \ref{fig:energy}. The receiving signal of the microphone is first segmented into frames and then converted to the frequency domain by short-time flourier transformation (STFT). Frequency selection strategy is used to select the frequency bins that are robust to the environment. Here, constant frequency bins from 3kHz to 6kHz are used for calculation which are proved to be stable by measurements in advance in our experiments. Other superior selection strategies can also be employed for calculation. For each frame, the energy with the selected bins denoted by ${U}$ are calculated and accumulated together. Thereafter, averaging is performed among all the frames, and this mean value multiplied by a coefficient $\beta$ (which is set as 0.8 in our experiments) is considered as threshold. The frames whose energy is greater than the threshold are reserved and averaged as the actual energy of the wakeup word. In this way, the silence and the noise portions are discarded, resulting in high accuracy of energy calculation.

\subsection{Calibration}
When competitive wakeup is enabled by APP, all the devices in the same WLAN response to the user competitively. At this momentum, when the user calls out the wakeup word, all the devices will receive the wakeup word and then calculate the energy of the wakeup word and that of the reference channel. This energy pair of one device is denoted by $\left( {{E_{mic,i}},{E_{spk,i}}} \right)$ (where $i$ is the device ID). When energy pairs of all the devices finished calculation, one device is randomly selected as master device. All the energy pairs together with their device ID which are denoted by $\left\{ {{\rm{ID}},\left( {{E_{mic,i}},{E_{spk,i}}} \right)} \right\}$ are transmitted to the master device for energy calibration and joint decision for selecting the responsor closest to the user. For example, if there are four devices in the WLAN, all the energy pair of the devices at the master device are denoted by $\left\{ {\# 1,\left( {{E_{mic,1}},{E_{spk,1}}} \right)} \right\}$, $\left\{ {\# 2,\left( {{E_{mic,2}},{E_{spk,2}}} \right)} \right\}$, $\left\{ {\# 3,\left( {{E_{mic,3}},{E_{spk,3}}} \right)} \right\}$, $\left\{ {\# 4,\left( {{E_{mic,4}},{E_{spk,4}}} \right)} \right\}$ respectively. The calibration matrix stored in the device is listed in Table~\ref{tab:calibration}.

Therefore, the actual energy of device can be calculated by

\begin{equation}
{E_i} = \underbrace {{b_i} \times {E_{{\rm{mic,}}i}}}_{\rm{microphone\ calibration}} - \underbrace {\sum\limits_{j = 1,j \ne i}^N {{a_{j,i}}{E_{{\rm{spk}},j}}} }_{\rm{speaker\ calibration}}
  \label{eq5}
\end{equation}

\begin{table}[H]
\centering
\caption{Calibration matrix.}
\label{tab:calibration}
 \begin{tabular}{|c|c|c|c|c|}
   \hline
   & $\#1$  & $\#2$  & $\#3$  & $\#4$  \\
   \hline
   $\#1$  & $b_1$  & $a_{1,2}$  & $a_{1,3}$ & $a_{1,4}$  \\
   \hline
   $\#2$  & $a_{2,1}$  & $b_2$  & $a_{2,3}$ & $a_{2,4}$  \\
   \hline
   $\#3$  & $a_{3,1}$  & $a_{3,2}$  & $b_3$ & $a_{3,4}$  \\
   \hline
   $\#4$  & $a_{4,1}$  & $a_{4,2}$  & $a_{4,3}$ & $b_4$  \\
   \hline
\end{tabular}
\end{table}

\subsection{Joint decision}
As long as the energy indicating the distance with respect to the user, and the DOA variance indicating the user orientation, the score can be combined by

\begin{equation}
{S_i} = \alpha  \cdot \left( {\frac{{{E_i}}}{{\sum\limits_{k = 1}^N {{E_k}} }} + \beta  \cdot \frac{{\frac{1}{{{G_i}}}}}{{\sum\limits_{k = 1}^N {\frac{1}{{{G_k}}}} }}} \right)
  \label{eq7}
\end{equation}
where, $S_i$ is the score, $E_i$ and $G_i$ are the energy and the DOA variance of the wakeup word calculated by eq.(\ref{eq6}) and eq.(\ref{eq5}), $N$ is the number of devices, $\beta$ is the proportion for orientation with respect to energy, $\alpha$ is the amplification factor. Since the score is a decimal, the $\alpha$ is used for amplifying the score for better comparison when implementation. These scores are sorted in the master, and the device with the maximal score can be determined. Therefore, a flag signal is transmitted to each device through WLAN to indicate the device to response or not.

\section{Performance Evaluation}
The influence factors for successful response caused by the environment are analyzed and validated by experiments.

\subsection{Network QoE}
\begin{figure}[H]
\centering
\includegraphics[scale=0.4]{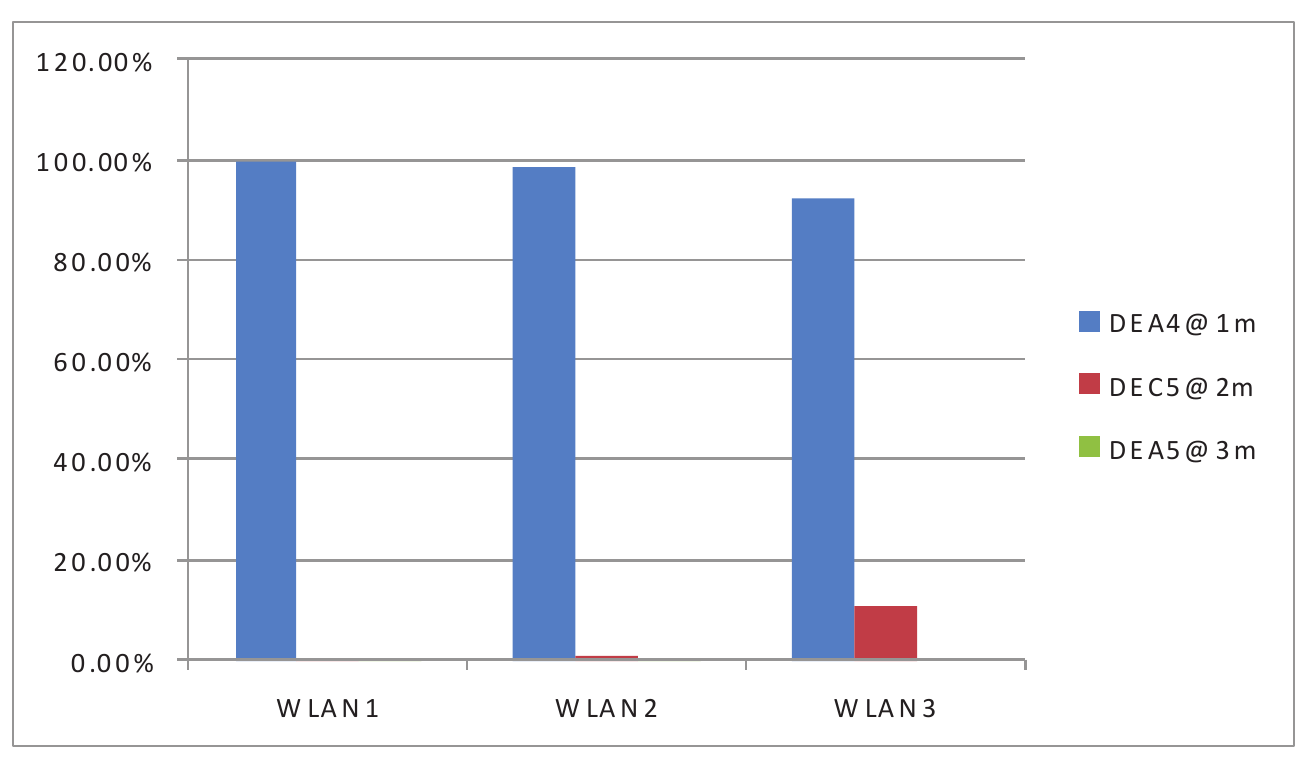}
\caption{Experiments at different network environments}
\label{fig:network}
\end{figure}

To verify the efficient of network QoE on the accuracy of competitive wakeup, three kinds of network are used for experiments, i.e., standalone router denoted by WLAN1, router with less connections denoted by WLAN2, router with plenty of connections denoted by WLAN3. The stability of these networks decreases one by one. Three devices labeled as DEA4, DEC5, DEA5 respectively are used for testing, which are locating at 1m, 2m, 3m with respect to the source in line, respectively. 200 corpus of wakeup word generated by five males and five females are used for playing at the source. The measurements are show in Fig. \ref{fig:network}. It can be found that it plays a key role of the QoE of network for keeping the accuracy of competitive wakeup.

\subsection{Voice orientation}
\begin{figure}[H]
\centering
\includegraphics[scale=0.3]{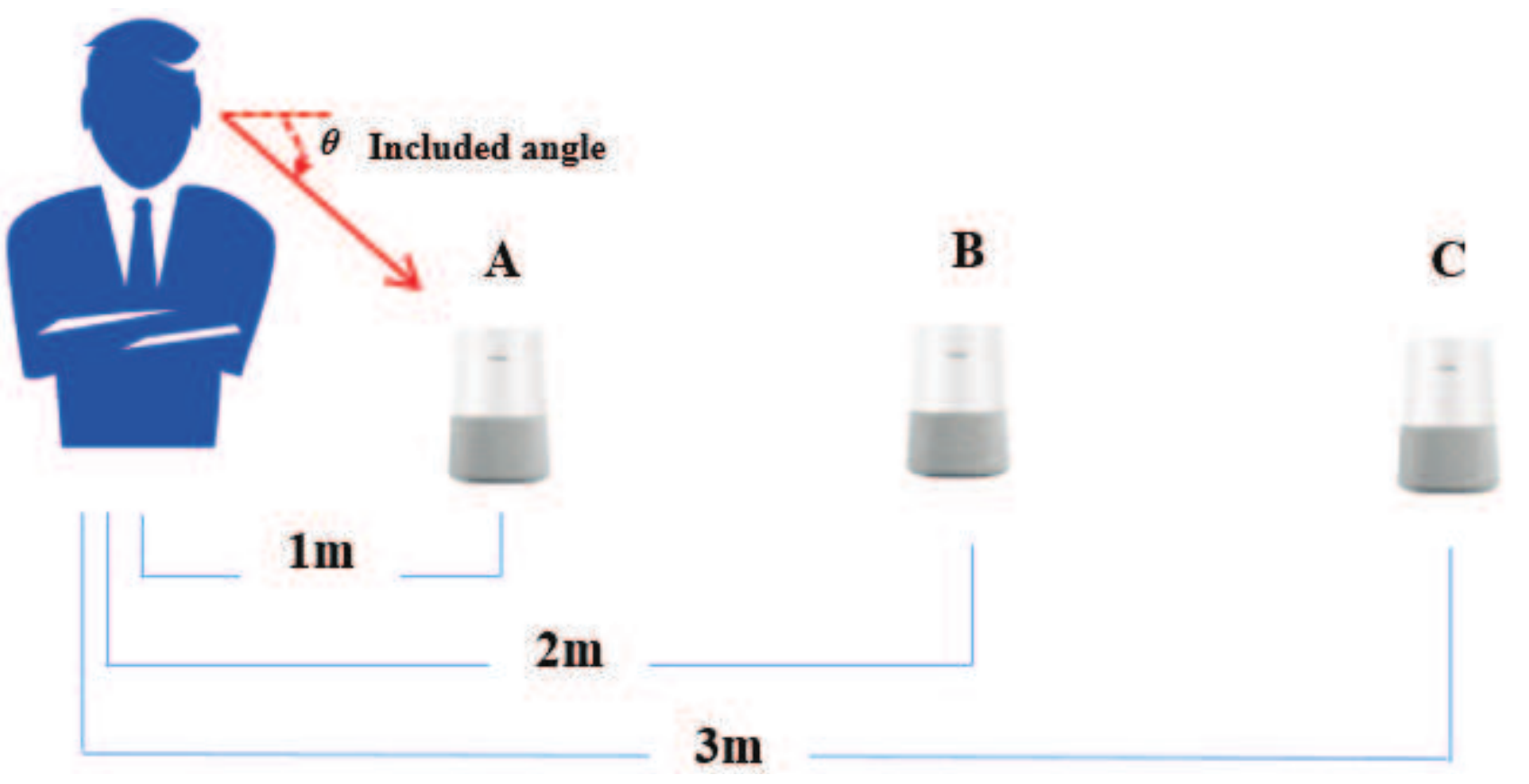}
\caption{Experiments deployment considering voice direction}
\label{fig:direction}
\end{figure}

Experiment is conducted in Fig. \ref{fig:direction} to analyze and validate the orientation performance. Two included angles are taking into considerations, i.e., $30^o$ and $60^o$.  Measurements reveal that the accuracy of $60^o$ recorded as 98$\%$ is higher than that of $30^o$ recorded as 85$\%$, i.e., the best method for waking up the corresponding device is to keep the eyes staring at the device.

\subsection{Ambient noise}
\begin{figure}[H]
\centering
\includegraphics[scale=0.5]{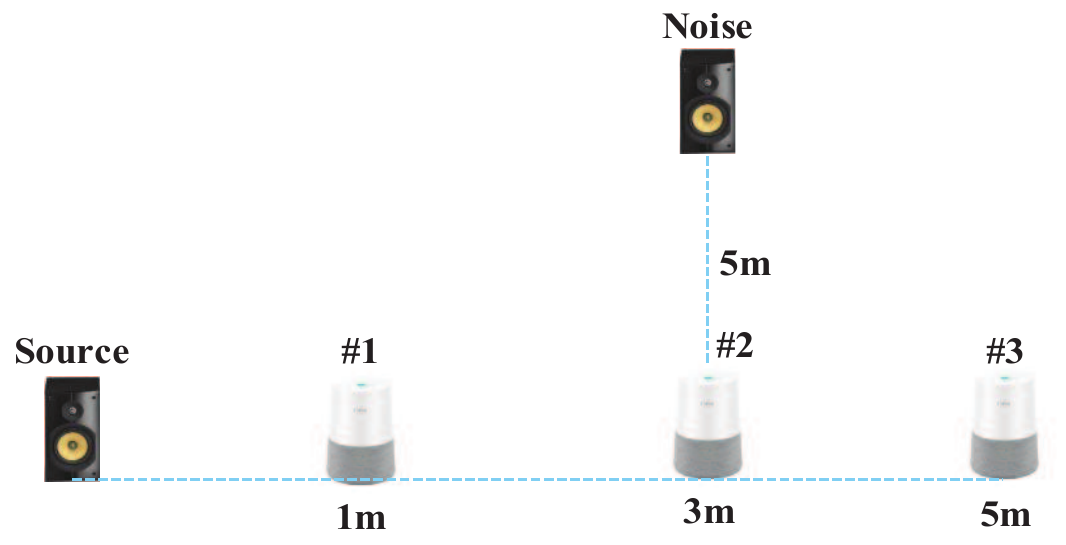}
\caption{Experiments deployment considering noise level}
\label{fig:noise}
\end{figure}

Experiments considering noisy environment should also be conducted to observe how the noise giving effect on the accuracy. The measurement deployment is show in Fig. \ref{fig:noise}, where the market noise recorded in advance from the real market environment is used to simulate this scenario. The noise level recorded at device $\#1$, device $\#2$ and device $\#3$ are 61dBa, 65dBa and 61dBa respectively, and the source level recorded at device $\#1$, device $\#2$ and device $\#3$ are 75dBa, 67dBa and 65dBa respectively. 200 corpus of wakeup word generated by five males and five females are used for playing at the source. The measured results are listed in Table~\ref{tab:accuracy}. It can be seen obvious that the noise will reduce the accuracy. This is not only because of the standalone failure of wakeup algorithm caused by noise, but also the inaccuracy of energy calculation introduced by noise.

\begin{table}[H]
\centering
\caption{Response accuracy}
  \label{tab:accuracy}
 \begin{tabular}{|c|c|c|c|c|c|}
   \hline
   Environment&  $1m$  & $3m$  & $5m$  & Failure & Accuracy  \\
   \hline
   Noisy  & 186  & 9  & 1 & 4 & $93\%$  \\
   \hline
   Quiet  & 198  & 1  & 1 & 0 & $99\%$  \\
   \hline
\end{tabular}
\end{table}

\section{Conclusions}
A competitive wakeup scheme is proposed to address the response problem when a same wakeup word is used for all the devices in the same network. The energy of the wakeup word received by the devices is employed for competition with elaborately designed calibration scheme, including the gain calibration of the microphone and the interference elimination introduced by the speakers. Experiments considering hardware diversity, network QoE, voice direction and environment noise are conducted and analyzed.

\bibliographystyle{IEEEtran}

\bibliography{mybib}


\end{document}